\documentclass[aps,prl,reprint,groupedaddress,showkeys,showpacs]{revtex4-1}
\usepackage{graphicx}
\usepackage{bm}
\usepackage{braket}

\begin{document}
\title{Excited-state potential-energy surfaces of metal-adsorbed organic molecules from Linear Expansion $\Delta$-Self-Consistent Field Density-Functional Theory ($\Delta$SCF-DFT)}

\author{Reinhard J. Maurer}
\email[]{reinhard.maurer@ch.tum.de}
\author{Karsten Reuter}
\email[]{karsten.reuter@ch.tum.de}
\affiliation{Department Chemie, Technische Universit{\"a}t M{\"u}nchen, Lichtenbergstr. 4, D-85747 Garching, Germany}

\date{\today}

\begin{abstract}
Accurate and efficient simulation of excited state properties is an important and much aspired cornerstone in the study of adsorbate dynamics on metal surfaces. To this end, the recently proposed linear expansion $\Delta$ Self-Consistent Field (le$\Delta$SCF) method by Gavnholt {\em et al.} [Phys. Rev. B {\bf 78}, 075441 (2008)] presents an efficient alternative to time consuming quasi-particle calculations. In this method the standard Kohn-Sham equations of Density-Functional Theory are solved with the constraint of a non-equilibrium occupation in a region of Hilbert-space resembling gas-phase orbitals of the adsorbate. In this work we discuss the applicability of this method for the excited-state dynamics of metal-surface mounted organic adsorbates, specifically in the context of molecular switching. We present necessary advancements to allow for a consistent quality description of excited-state potential-energy surfaces (PESs), and illustrate the concept with the application to Azobenzene adsorbed on Ag(
111) and Au(111) surfaces. We find that the explicit inclusion of substrate electronic states modifies the topologies of intra-molecular excited-state PESs of the molecule due to image charge and hybridization effects. While the molecule in gas phase shows a clear energetic separation of resonances that induce isomerization and backreaction, the surface-adsorbed molecule does not. The concomitant possibly simultaneous induction of both processes would lead to a significantly reduced switching efficiency of such a mechanism.
\end{abstract}



\maketitle 

\section{Introduction}
\label{intro}

Functional organic adsorbates are of special interest to surface nanotechnology. Their ability to selectively trigger dynamical changes in surface domains on molecular length scales opens many relevant applications in technology. A specifically interesting class of functional molecules are molecular switches\cite{Ferringa2001}, which can be reversibly switched between two or more stable geometries. Switching of these molecules adsorbed on surfaces can occur via photo-excitation\cite{Comstock07,Comstock2008,Schmidt2010,Morgenstern2011} or inelastic electron scattering events\cite{Comstock2005,Choi2006,Alemani2008}. A typical problem to the design of such systems is the loss of photo-induced switching function of gas-phase molecules upon adsorption to a surface, such as in the case of the prototypical conformational cis-trans (Z-E) switch Azobenzene (Ab) on coinage metal surfaces\cite{Bronner2012}. Very often this loss of function is related to overly strong coupling to the surface electronic degrees of 
freedom. This coupling modifies ground-state energetics\cite{Maurer2012} and, even more importantly, is expected to change excited-state mechanisms and lifetimes. For the Ab case recent experiments have e.g. shown that for the supposedly minimized coupling of a derivative of Azobenzene with bulky spacer groups, namely \textit{tetr}-tert.-butyl-Azobenzene (TBA), the switching function can indeed be retrieved at Au(111) \cite{Comstock07,Schmidt2010}, while no switching is observed for the azobenzene and TBA on Ag(111).

A plethora of spectroscopical techniques gives access to the changes in electronic structure that underlie corresponding excited-state dynamical processes. Still, often enough the accessible observables do not permit to formulate a precise molecular mechanism, which in turn is a prerequisite to the understanding and subsequent design of corresponding systems. First-principles modelling techniques have in turn proven to be valuable tools for the investigation of such mechanistic details, but are challenged by the large system sizes and the necessity to simultaneously describe localized Molecular Orbitals (MOs) and the metallic surface band structure. {\em Ab initio} quantum mechanical simulations, such as Density-Functional Theory (DFT)\cite{Hohenberg1964,Kohn1965} or post-Hartree-Fock approaches\cite{Szabo1989}, have a successful history as such tools in surface science and chemistry. The current state-of-the-art provides, in many cases, a reliable description of ground-state properties, including adsorption 
geometries, energetics as well as thermal barriers. When it comes to the description of spectroscopy and excited-state properties, quantum chemical approaches are the optimal choice for finite systems or isolating materials, where cluster approximations are possible. They are currently not applicable to metallic systems though, where periodic boundary conditions are necessary to correctly describe the delocalized electronic structure. Applicable excited-state methods for this case include Time Dependent DFT (TD-DFT) \cite{Runge1984,Onida2002,Casida2009}, or many-body perturbation theory (MBPT) based methods\cite{Hedin1965}, the latter enabling the description of both, ionic (GW)\cite{Aryasetiawan1998} and neutral (Bethe-Salpeter equation, BSE)\cite{Sham1966,Hanke1975,Onida2002} electronic excitations. In recent years, computational cost and accuracy of these approaches has improved tremendously. Nevertheless, current computer infrastructure and the remaining accuracy issues of applied density-functional 
approximations and self-energy descriptions render systematic excited-state studies of large systems virtually intractable at the time. In this situation, a need exists for highly efficient excited-state schemes that, while maybe not fully quantitative, allow for a qualitatively correct description of the major physical effects that govern excited-state processes at surfaces.

Already very early in the history of DFT, attempts to apply and/or generalize the method beyond Hohenberg and Kohns rigorous proof\cite{Hohenberg1964} and towards non-equilibrium excited-state properties have been undertaken. Many of them with the specific aim for a highly efficient description. The most rigorous and major extension was the Runge-Gross proof of a one-to-one correspondence between the time-dependent potential and the time-dependent electron density\cite{Runge1984}. Another line of development are functionals generalized to fractional occupation numbers\cite{Gross1988,Weinert1992}, which has led to the standard DFT treatment for metallic systems\cite{Kresse1996}. Utilizing a Lagrange multiplier formalism, Dederichs \textit{et al.}\cite{Dederichs1984} have shown how to construct constrained density functionals\cite{Kaduk2011}, constraining electrons into specific regions of space or spin channels. This very efficient method has been heavily utilized to describe electron-transfer processes\cite{
Wu2006, Oberhofer2009, Oberhofer2010}, but also surface reactions\cite{Behler2007a, Behler2008}. Another very early approach is based on converging the density with respect to non-equilibrium electron occupations that resemble excitations, so called Delta-Self-Consistent Field DFT ($\Delta$SCF-DFT)\cite{Clementi1962,Gunnarsson1976,Ziegler1977}. This approach, in different variations, has had a comeback in recent years due to its success on molecular charge-transfer excitations\cite{Hellman2004,Baruah2009, Kowalczyk2011, Maurer2011,Kasamatsu2011, Baruah2012, Himmetoglu2012}, which are badly described by adiabadic linear-response (lr) TD-DFT using standard semi-local exchange-correlation (xc) kernels\cite{Peach2008, Wiggins2009}. Although in principle without any formal justification, this method has recently been put into context by a number of different works. Ziegler and coworkers identified a close connection to a constrained variational procedure\cite{Ziegler2009,Cullen2011,Ziegler2011,Ziegler2012}, which 
then provides a direct link to lrTD-DFT\cite{Cullen2011}. Theoretical works by G\"orling\cite{Gorling1999} and Ayers \textit{et al.}\cite{Ayers2009} in turn point towards a possible formal basis for an excited-state density functional and would, at least in the case of the exact xc-functional, justify a corresponding treatment.

In the context of metal-surface adsorbed molecules an interesting extension to $\Delta$SCF-DFT was put forward by Gavnholt \textit{et al.}\cite{Gavnholt2008}. This so-called linear expansion $\Delta$SCF (le$\Delta$SCF) scheme centers on resonance states that resemble gas-phase adsorbate orbitals, and enforces their occupation in the self-consistent density. This not only provides a well-defined constraint for intra-molecular excitations of the adsorbate, but also enables the description of photoemission and charge-transfer excitations. The method has already been successfully applied to several smaller adsorbate systems\cite{Olsen2009,Zawadzki2011,Garcia-Lastra2011} and promises at least semi-quantitative results, while adding only little computational overhead to ground-state DFT calculations. In this work we strive to generalize this le$\Delta$SCF approach as an efficient means to the calculation of excited-state potential-energy surfaces (PESs) for large metal-adsorbed molecules. We present necessary 
modifications to the method that allow the calculation of intra-adsorbate as well as substrate-mediated charge-transfer excitations within the same formalism. Implementing the method into the ultrasoft pseudopotential plane wave code CASTEP\cite{Clark2005}, we initially compare the ability of this le$\Delta$SCF scheme to describe excited-state PESs in the gas phase, and establish the equivalence to simple $\Delta$SCF for this limit of no molecule-substrate interaction. Thereafter we illustrate what kind of insights can be gained from this approach for the showcase Azobenzene adsorbed on extended Ag(111) and Au(111) surfaces. Already from the computed lowest-lying intra-molecular excitations along the two most important gas-phase isomerization pathways we can conclude on intriguing surface-induced PES modifications with direct bearings on the isomerization mechanism.

\section{Methods}
\label{theory}

In the following section we briefly revisit the $\Delta$SCF method and introduce the rationale behind le$\Delta$SCF as well as our modification and implementation of it. 

\subsection{$\Delta$Self-Consistent-Field-DFT and le$\Delta$SCF-DFT}
\label{ledscf}
                                         
Detailed descriptions of the $\Delta$SCF-DFT approach have already been given numerous times in literature\cite{Gunnarsson1976,Hellman2004,Maurer2011,Kowalczyk2011,Baruah2012}. In the simplest case, excitations are modelled by reordering orbital occupation between states that mainly contribute to a transition. This changed population generates a modified density under which the Kohn-Sham (KS) equations\cite{Kohn1965} are solved. Singlet excitations are modelled by changing populations within one spin channel, Triplet excitations by switching channels. Certain care has to be taken to ensure the correct calculation of Singlet states. A generally used correction method is Zieglers sum rule\cite{Ziegler1977}: $E^{SM}_S=2E_S-E_T$. If the system does not show magnetization and the ground state is a Singlet state, Singlet excitations can also be calculated to a reasonable approximation without taking spin explicitly into account. This has been shown for the case of gas phase Azobenzene\cite{Maurer2011}, O$_2$ on Al(
111)\cite{Behler2007a} and recently also for Iridium complexes\cite{Himmetoglu2012}, the latter work also providing a rationalization for the success of this approach. 

Corresponding constraints provide a reasonable description of excited states that are well described as single-particle state-to-state transitions, although the variational adaptation of the KS states with respect to the excited-state density clearly does give additional flexibility. Definition of such single-electron excitation constraints is a simple matter when molecular states can be clearly identified in character and are well separated spatially and energetically. This is almost always the case in minimum-energy structures of isolated organic molecules. More reactive geometries, \textit{i.e.} transition-state structures, can already contain state degeneracies that hamper convergence. In such a case minimal smearing of the occupation constraint might enable calculation with only a small additional error in energy\cite{Maurer2011}. In contrast, in the case of the excitation spectrum of molecules interacting with periodic structures, where degeneracies are ubiquitous, such a simple approach will strongly 
affect the character and the absolute energy of the excitation. In this situation one also has to distinguish between adsorbates interacting with isolating surfaces and adsorbates on metals. In the first case, substrate states are mainly localized and generally exhibit strong hybridization with adsorbate states similar to interactions between two covalently interacting molecules. Such hybridization can in principle completely modify the character and the energy of states, but will again generate states that are localized and can, in the best case, be identified in their character and occupied correspondingly. Therefore a simple $\Delta$SCF approach should still capture the main part of the transition. In the case of transition metal substrates, however, interactions are twofold. Following the Newns-Anderson model\cite{Gross}, chemisorbed molecules will show strong hybridization with \textit{d}-bands, which modifies character, splitting and energetic position of the frontier orbitals. Simultaneously, there 
will also be a weak hybridization due to interaction with \textit{s}- and \textit{p}-bands. This broadens molecular states and spreads their character over many bands in a small energy window. In such a case a simple $\Delta$SCF approach, that occupies the band with the highest overlap compared to a gas-phase molecular state, will miss significant parts of the transition and therefore strongly underestimate the change in density.

Gavnholt \textit{et al.}\cite{Gavnholt2008} have devised the linear expansion $\Delta$SCF (le$\Delta$SCF) approach to specifically target such systems. In their approach they do not just define constraints on KS states, but on linear combinations of them. To illustrate this, let us shortly recapitulate the ground-state case for an isolated system (or for an extended system for each $\mathbf{k}$-point separately). There the effective one-particle KS equations read 

\begin{equation}
 \left[ -\frac{\nabla^2}{2}+V_{\text{KS}}[\rho] \right] \ket{\psi_i} = \epsilon_i \ket{\psi_i}
\end{equation}
where we define the KS potential V$_{\text{KS}}$ acting on the KS auxiliary wavefunctions and the KS eigenvalues $\epsilon_i$. The density (in the following written in Dirac notation) on which the KS potential depends on, is constructed from the \{$\ket{\psi_i}$\} via
\begin{equation}
 \rho=\sum_{i}^{\text{states}} f_i \ket{\psi_i}\bra{\psi_i} \quad ,
\end{equation}
where $f_i$ is the occupation of the state i. In a $T = 0K$ ground-state calculation this results in 
\begin{equation}
 \rho=\sum_{i=1}^{N_e} \ket{\psi_i}\bra{\psi_i}
\end{equation}
for a finite system with $N_e$ being the number of electrons of the system, or in case of an extended system
\begin{equation}
 \rho=\sum_\mathbf{k} w^{\mathbf{k}}\sum_{i}f_i(\epsilon_F) \ket{\psi_{i}^{\mathbf{k}}}\bra{\psi_{i}^{\mathbf{k}}}
\end{equation}
with $w^\mathbf{k}$ being the mathematical weight for each $\mathbf{k}$-point and $\epsilon_F$ being the Fermi energy. In simple $\Delta$SCF calculations one instead constructs the density by replacing one of the states in the sum with another originally unoccupied virtual KS state. In le$\Delta$SCF, Gavnholt \textit{et al.} propose to construct so-called resonance states from a linear combination of KS states instead of a single KS state
\begin{equation}\label{ledscfpsi}
 \ket{\tilde{\psi}_{c}^\mathbf{k}}=\sum_{i}^{\text{unocc.}} a_i^\mathbf{k}\ket{\psi_{i}^\mathbf{k}}
\end{equation}
with expansion coefficients $a_i^\mathbf{k}$ defined as
\begin{equation}
 a_i^\mathbf{k}=\frac{\braket{\psi_i^\mathbf{k}|\phi_c^\mathbf{k}}}{(\sum_i|\braket{\psi_i^\mathbf{k}|\phi_c^\mathbf{k}}|^2)^{1/2}} \quad ,
\end{equation}
where $\ket{\phi_c^\mathbf{k}}$ denotes a pre-calculated reference KS state of the corresponding gas-phase adsorbate that ought to be occupied. The excited-state density then follows as
\begin{equation}\label{ledscfdens}
 \rho=\sum_\mathbf{k}w^{\mathbf{k}} \left( \sum_{i=1}^{\text{occ.}} \ket{\psi_i^\mathbf{k}}\bra{\psi_i^\mathbf{k}} + \sum_{i,j}^{\text{unocc.}}a_i^{\mathbf{k}}\cdot a_j^{\mathbf{k}*}\ket{\psi_i^\mathbf{k}}\bra{\psi_j^\mathbf{k}} \right) \quad .
\end{equation}

Equation \ref{ledscfpsi} thus constructs a new KS state from unoccupied orbitals which resemble the chosen reference state and which are then used to construct the excited-state density. This approach can readily be used to model intra-molecular highest occupied MO (HOMO)- lowest unoccupied MO (LUMO) type excitations, where an equal number of electrons and holes are excited in the adsorbate, but also for adsorbate-substrate charge-transfer, where only a hole or an additional electron is enforced on the adsorbate states. In this case, the occupation of the remaining states has to be adjusted by lowering or increasing the Fermi energy correspondingly in order to conserve the total electron number of the whole system. This approximation to an excitation can be justified by the large ensemble of substrate electrons which occupy metal bands of very similar character, such that removing one such band from the density should induce only a minor error on the excitation energy. Summarizing the approach, those parts 
of the reference orbital, which are not yet included in the first N$_e$-1 ground state occupied orbitals are constructed from a range of virtual KS states by projecting out components resembling this state and subsequently including them in the density. The ground-state KS procedure is therefore modified in the construction of the density (eq. \ref{ledscfdens}) in every SCF step. Following this approach the kinetic energy of the system has to be corrected for the terms due to the newly added constraint orbital\cite{Gavnholt2008}. When breaking spin symmetry or including different positions in momentum space ($\mathbf{k}$-point sampling) the procedure is followed independently for different spin channels or at different $\mathbf{k}$-points. This approach is ideally suited for the description of inverse photo-emission and for diatomics on transition metal surfaces and was shown to outperform spatially constrained DFT approaches as well as simple $\Delta$SCF \cite{Gavnholt2008}.

\subsection{A fresh look on le$\Delta$SCF}
\label{ledscf-improve}

In the following we would like to generalize the le$\Delta$SCF approach in two aspects. First of all to allow for an arbitrary number of constraints constructed from arbitrary reference states without discriminating between occupied and unoccupied states. This provides a more consistent infrastructure for the description of intra-molecular as well as charge-transfer excitations, and might even open the application to systems very different from adsorbate-substrate complexes. Secondly, the approach should enable the construction of excited-state PESs for large adsorbates in arbitrary geometries, while in the limit of infinite separation between adsorbate and substrate it should retrieve the simple $\Delta$SCF result. In order to achieve this, certain conditions on the reference states $\ket{\phi_c}$ have to be imposed.

\paragraph{Modified Approach}

It is always possible to expand an arbitrary reference state $\ket{\phi_c}$ in the complete space of KS states of the system under study,
\begin{equation}\label{leDSCFeq}
 \ket{\tilde{\psi}_c^{\mathbf{k}}}=\sum_i^{\text{states}} \ket{\psi_i^{\mathbf{k}}}\braket{\psi_i^{\mathbf{k}}|\phi_c^{\mathbf{k}}},
\end{equation}
while at the same time the remaining KS states have to be orthogonalized correspondingly
\begin{equation}
 \ket{\tilde{\psi_i^{\mathbf{k}}}}=\ket{\psi_i^{\mathbf{k}}}-\sum_c^{\text{constr.}}\ket{\phi_c^{\mathbf{k}}}\braket{\phi_c^{\mathbf{k}}|\psi_i^{\mathbf{k}}} \quad .
\end{equation}
This leaves the subset of \{$\ket{\tilde{\psi_i^{\mathbf{k}}}}$\} orthogonal to the subset of resonance states \{$\ket{\tilde{\psi_c^{\mathbf{k}}}}$\}, but destroys orthonormality for the complete set of KS states \{$\ket{\tilde{\psi_i^{\mathbf{k}}}},\ket{\tilde{\psi_c^{\mathbf{k}}}}$\}. We therefore perform an additional orthonormalization on this whole set of KS states. This state transformation is done in every SCF step and yields a set of KS states on which a simple modification of the electron occupation, such as it is done in simple $\Delta$SCF, yields an excited-state density as follows
\begin{equation}\label{projdscfdens}
 \rho^{'} =\sum_\mathbf{k}w^{\mathbf{k}} \left( \sum^{\text{states}\neq\text{constr.}}_i f_i^{'} \ket{\tilde{\psi_i^{\mathbf{k}}}}\bra{ \tilde{\psi}_i^{\mathbf{k}}} +\sum^{\text{constr.}}_{c} f_c \ket{\tilde{\psi}_c^{\mathbf{k}}}\bra{\tilde{\psi}_c^{\mathbf{k}}} \right) \quad ,
\end{equation}
where the only boundary condition on eq. \ref{projdscfdens} is that
\begin{equation}\label{occ_constraint}
 \sum^{\text{states}\neq\text{constr.}}_i f_i^{'}+\sum_c^{\text{constr.}}f_c=N_e \quad .
\end{equation}
In eq. \ref{occ_constraint} the $f_c$'s are given by the aspired constraint definition, while occupations $f_i'$ have to be adapted to conserve the electron number. Due to the modified construction of the KS states and the occupation reordering, there is no need for modification of the density construction routine itself. The modified KS states and excited-state occupations also enter in the calculation of the kinetic energy, which is therefore implicitly treated correctly, again without need of further modification as was necessary in the original implementation of Gavnholt \textit{et al.} in the GPAW package\cite{Mortensen2005,Enkovaara2010a}. Constraints are enforced independently in different spin channels and at different $\mathbf{k}$-space positions.

This approach only differs from the simple $\Delta$SCF approach by the modification of the KS states, which corresponds to a unitary transformation and forces the resulting KS solution to include the specified resonances. It naturally accounts for the hybridization-induced broadening of the adsorbate KS states at the surface and in all cases includes or removes the \textit{whole} reference state. In contrast, a strong limitation of the method lies in hybridization effects that go beyond this. Due to the interaction of the sub-systems (molecule and surface), hybridization of the system can already lead to ground-state occupations that are very different from the separated sub-system case. The correct treatment of electron transitions then has to start from this occupation and transfer the corresponding amount of electrons effectively as we will discuss in the application to adsorbed Azobenzene below. 

Concerning the calculation of energy derivatives, this approach suffers from similar problems as the original implementation of le$\Delta$SCF does. The Hellmann-Feynman theorem does not hold due to the additional dependence of the non-variational coefficients of $\ket{\phi_c}$ on the positions of nuclei and the additional entropic contribution due to the excited-state population. A possible formulation of the herewith introduced $\ket{\phi_c}$-\lq Pulay\rq-like terms still needs to be developed.

\paragraph{Generating Suitable Reference States}

\begin{figure}
\includegraphics[width=8cm]{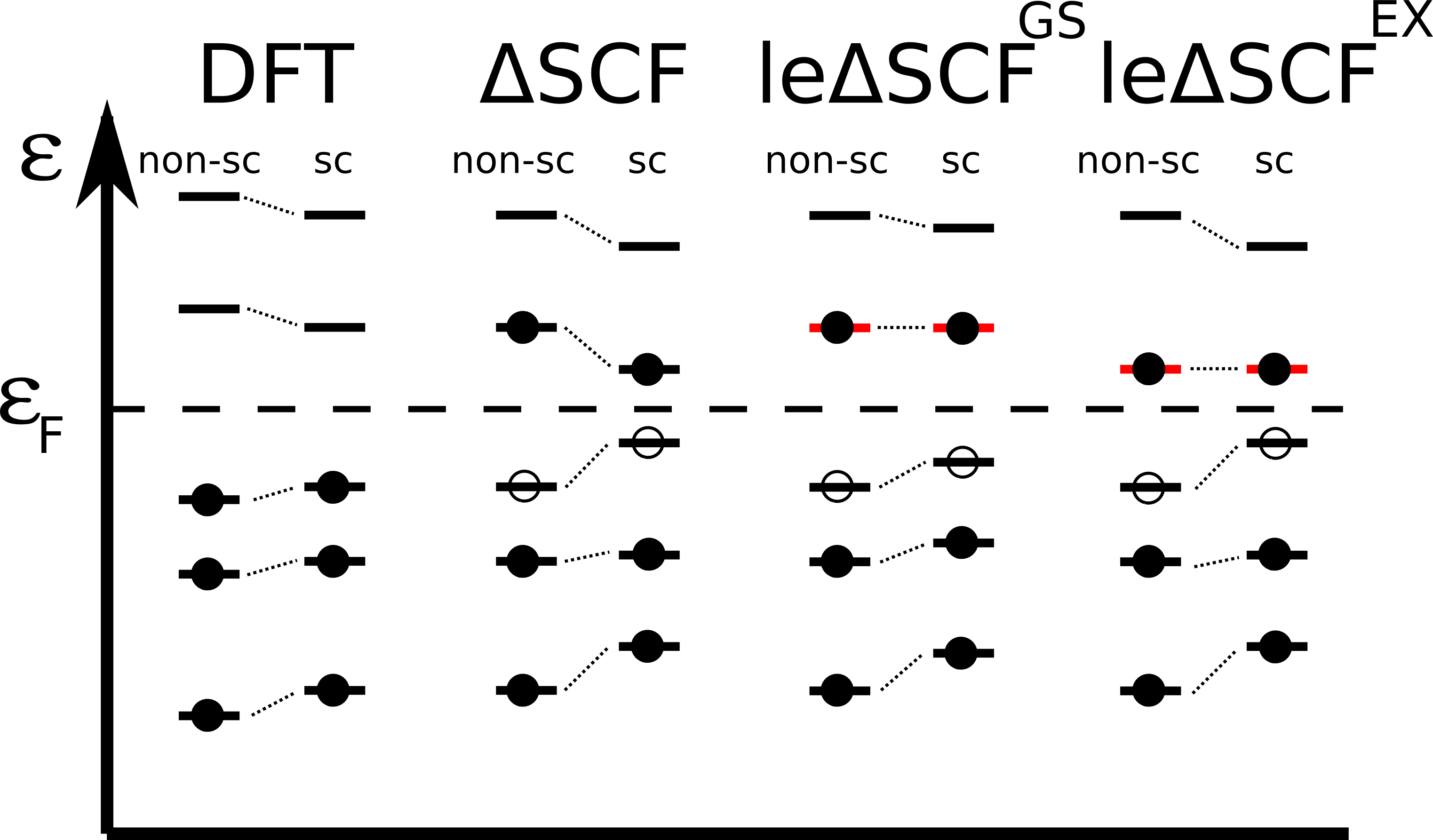}
\caption{\label{dscf-scheme} Schematic state diagram showing the frontier orbitals in ground-state DFT, simple $\Delta$SCF, le$\Delta$SCF with a ground-state reference and le$\Delta$SCF with an excited-state reference. Projected reference states are in red. The excitation is visualized with an electron hole pair (filled and unfilled circles). Schematic orbital positions are shown for the non-self-consistent (non-sc) and the self-consistent case (sc).}
\end{figure}

In the le$\Delta$SCF scheme an excited-state density is constructed that includes a certain resonance state. All remaining states are variationally relaxed and therefore effectively screen the excitation in the self-consistent (sc) excited-state density. A question that remains is the selection of suitable reference states $\ket{\phi_c}$ from which to construct the resonances. Such reference states could be molecular states of an adsorbate on a surface resembling an excitation, localized orbitals of a cluster cut-out that resemble a vacancy, or stemming from the very same system in a different electronic state (as used in the case of Ab in gas phase below). The choice depends very much on the definition of the sub-system and the excitation under study. The projection restricts the resonance state itself to be an input quantity and it cannot change during the self-consistent solution of the KS equations. This stands in stark contrast to simple $\Delta$SCF where the 
non-self-consistent (non-sc) input orbitals from the ground-state calculation are used to construct the input density and are then iteratively optimized to yield a self-consistent excited-state density($\Delta$SCF, \textit{cf.} Fig. \ref{dscf-scheme}). 

In the work of Gavnholt \textit{et al.} the choice of the reference state fell on a virtual ground-state KS state of the gas-phase adsorbate ($\Delta$SCF$^{\text{GS}}$). As shown schematically in column 3 of Fig. \ref{dscf-scheme}, this corresponds to calculating the sc excited-state solution, while forcing the constrained orbital into the non-sc (or ground-state) solution. This might be a valid approximation, if the molecular state of interest does not change strongly due to screening in the excited state. Particularly for the description of vertical excitation energies of equilibrium geometries or PESs of small adsorbates where a small number of degrees of freedom defines the KS states, this might be a good choice. This was nicely shown for the calculation of excited-state PESs of small diatomics on transition metal surfaces\cite{Gavnholt2008,Olsen2009}. The single nuclear degree of freedom in this case reduces the chance of large variations in character and extent of the orbitals between ground-state and 
excited-state solutions. 

Notwithstanding, in many cases this approximation will fail, namely when the ground-state optimized orbital is not a good approximation to the final excited-state KS state. This is in fact the general case for the frontier orbitals of molecules with many degrees of freedom and/or extended $\pi$-systems in non-equilibrium geometries, and is especially true when applying standard semi-local exchange-correlation (xc) functionals. In the latter case it is known from lrTD-DFT treatments that the qualitatively wrong description of ground-state molecular resonances in semi-local functionals hampers the description of non-equilibrium geometries and charge-transfer excitations\cite{Ploetner2010,Dwyer2010, Maurer2011}. This problem can to some extent be resolved by including the correct 1/r density-density-response behaviour into the xc-functional description\cite{Tozer1998a}. A big strength of the simple $\Delta$SCF approach is in this respect its additional flexibility due to the variational optimization of the 
orbitals. Although the definition of the excitation itself is primitive compared to lrTD-DFT, the additional variation allows for a consistent-quality description for large portions of PESs and the qualitatively correct description of charge-transfer states and other problematic cases already with a semi-local or hybrid xc-functional\cite{Maurer2011,Kowalczyk2011,Baruah2012, Himmetoglu2012}. The absolute excitation energies will nonetheless be determined by the quality of the underlying xc functional, meaning that an underestimation of \textit{e.g.} the HOMO-LUMO gap due to self-interaction error will also carry over to the excited-state description.

Some of this $\Delta$SCF flexibility is lost due to the projection inherent in le$\Delta$SCF. In order to also ensure a correct sc treatment of the actually constraint orbitals, one has to provide reference states $\ket{\phi_c}$ that are already optimized to the specific excited state of interest. This can for example be done by calculating the simple $\Delta$SCF solution of the excited-state reference system (here the gas-phase molecule) and then including reference states into the le$\Delta$SCF calculation that are already in the final excited state (le$\Delta$SCF$^{\text{EX}}$, cf. Fig. \ref{dscf-scheme}). The solution of this approach would in this case correspond to the simple $\Delta$SCF solution in the limit of zero hybridization. Such a generalized final-state le$\Delta$SCF (le$\Delta$SCF$^{\text{EX}}$) approach to arbitrary systems would thus include the following steps:

\begin{itemize}
 \item Calculate the electronic ground-state of the system of interest with DFT
 \item Calculate the excited state of interest in the reference (sub-)system using simple $\Delta$SCF-DFT
 \item Calculate the excited state of the system of interest using the excited reference state and le$\Delta$SCF-DFT
\end{itemize}

\subsection{Computational Details}
\label{methods}

The method described above has been implemented in the ultrasoft pseudopotential plane-wave code CASTEP 6.0.1\cite{Clark2005}. The implementation for the $\Delta$SCF scheme constructs the changed set of KS states after every diagonalization step in the SCF procedure and uses a modified Fermi distribution to assign the constraint occupations, adapt the remaining occupations ($f_i'$) and construct the density from it. The newly constructed resonance KS state replaces the former KS state showing the highest overlap with the reference state. Calculations employing the le$\Delta$SCF method as implemented in CASTEP need to be checked for convergence with respect to the standard parameters of plane wave calculations such as plane wave cutoff and k-point sampling, but also with respect to the number of additional virtual orbitals that are explicitly included in the calculation in order to assure convergence of the projections from eq. \ref{leDSCFeq}. Standard DFT convergence enhancement 
methods\cite{Payne1992, Kresse1996, Hasnip2006} are used for the evaluation of the self-consistent density. In addition to the modified $\Delta$SCF scheme, simple $\Delta$SCF for gas-phase molecules in a supercell approach has been implemented. This is used for comparison and construction of appropriate excited-state KS reference states. The implementation of the projections in eq. \ref{leDSCFeq} allowed us to use them also for the calculation of Molecular Orbital projected Density-of-States (MolPDOS) following the explanations of McNellis {\em et al.} \cite{McNellis2009a}. MolPDOS coefficients corresponding to gas-phase reference KS states can be printed out and post-processed for visualization (as shown in ref. \onlinecite{Maurer2012}). These coefficients give access to the MO occupations that are shown in Fig. 3 and 4.

Excitations in this work have been modelled by effective addition or removal of one electron in the frontier molecular orbitals, namely, the second highest occupied molecular orbital (HOMO-1), HOMO and LUMO of the molecule in order to describe neutral intra-molecular excitations.

During this work we have used standard-library ultrasoft pseudopotentials (USPPs)\cite{Vanderbilt1990}. We use the xc-functional due to Perdew, Burke and Enzerhof (PBE) \cite{Perdew1996} throughout. Isolated Ab benchmark calculations below have been run in a 20x20x20 \AA\ supercell with a plane-wave energy cut-off of 350\,eV and $\Gamma$-point sampling. The corresponding geometries along the gas-phase isomerization pathways have been taken from a previous study\cite{Maurer2011}. Metal-surface adsorbed Ab calculations were done in (111) (3x6) 4-layer surface slabs of Ag(111) and Au(111), employing a plane wave cutoff of 350\,eV and 16 irreducible k-points. The vacuum region was chosen to exceed 20\,\AA. This calculational setup closely follows refs \onlinecite{Maurer2012} and \onlinecite{McNellis2009}, where careful convergence tests have already been detailed. Optimized structures for the minimum-energy paths at the surface were taken from a recent work\cite{Maurer2012}. A semi-
empirical dispersion correction was employed to ensure correct description of the adsorbate geometry\cite{Tkatchenko2009,McNellis2009}, although explicitly excluding lateral interactions in order to describe the system in a low-coverage limit.

\section{Results and Discussion}
\label{results}

In this section we apply the proposed method first to the isomerization of the prototypical molecular switch Azobenzene in gas phase and then when adsorbed on coinage metal surfaces.

\subsection{Isomerization of gas-phase azobenzene}
\label{gasphase-azo}
 \begin{figure*}
 \includegraphics[width=\linewidth]{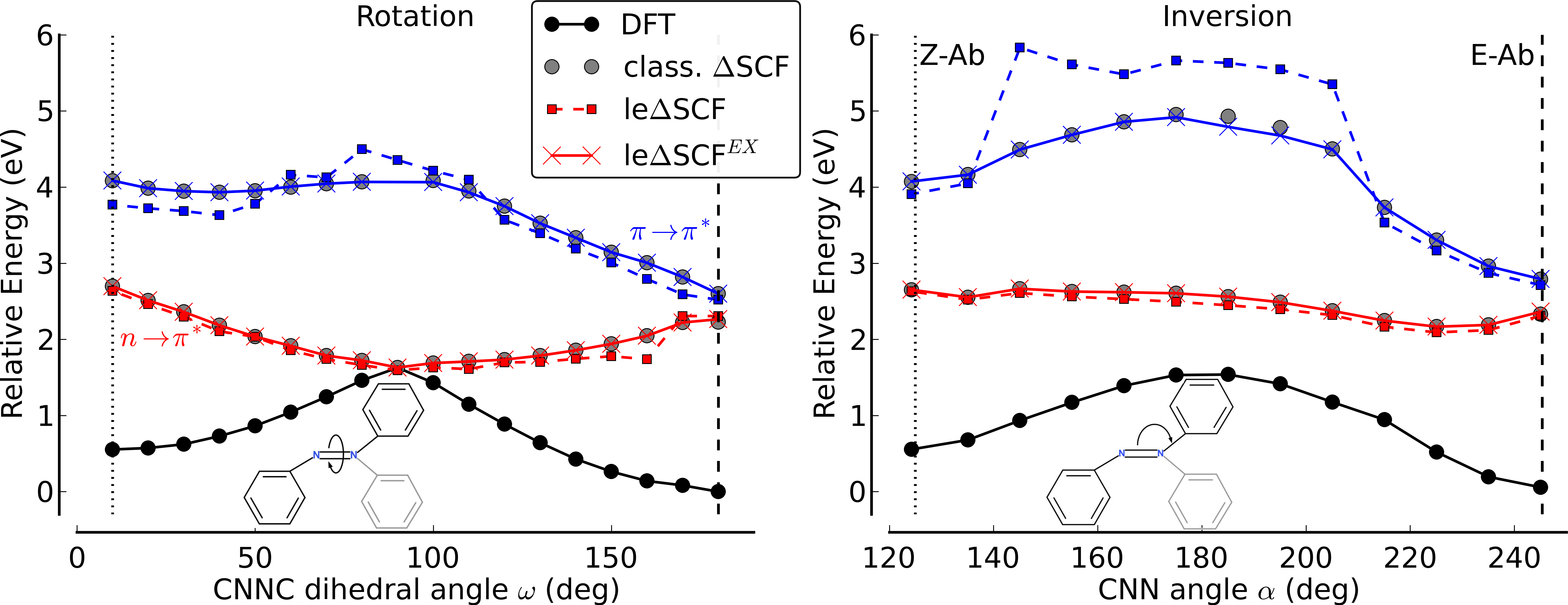}
 \caption{\label{refchange} Gas-phase Azobenzene PESs along rotation (left) and inversion (right) degrees of freedom for the ground state (black), the first excited n$\rightarrow\pi^*$ state (red) and the second excited $\pi\rightarrow\pi^*$ state (blue). The excited-state curves were calculated with simple $\Delta$SCF (no lines, gray circles), le$\Delta$SCF with ground-state reference orbitals (dashed lines, squares) and le$\Delta$SCF with excited-state reference orbitals (straight lines, crosses). The insets illustrate the corresponding Azobenzene degrees of freedom of dihedral rotation $\omega$ and inversion along one CNN angle $\alpha$.}
 \end{figure*}

The purpose of this subsection is to demonstrate the equivalence of simple $\Delta$SCF and le$\Delta$SCF in the limit of photo-induced E-Z-isomerization of gas-phase Azobenzene. Azobenzene can efficiently Z$\rightarrow$E or E$\rightarrow$Z isomerize upon UV-excitation to the S1($\text{n}\rightarrow\pi^*$) and S2($\pi\rightarrow\pi^*$) state, respectively\cite{Rau2003}. Several recent experimental\cite{Nagele1997, Fujino2002, Schultz2003, Satzger2003, Satzger2004} and theoretical works\cite{Cattaneo1999, Schultz2003, Cembran2004, Conti2008, Yuan2008, Ootani2009, Bockmann2010} suggest that the mechanisms for both directions of isomerization mainly follow S1 dynamics, due to strong initial population transfer from the S2 state to the S1. This dominant S1 dynamics is very often discussed in terms of two mechanisms, namely, rotation around the central CNNC dihedral angle $\omega$ and inversion around one of the two NNC angles $\alpha$ (\textit{cf.} insets in Fig. \ref{refchange}), and there is a long-lasting 
controversy as to the prevalence of one of these two mechanisms\cite{Rau1982, Rau1984}. Several recent \textit{ab initio} non-adiabatic dynamics studies have come to the common conclusion though that the dominant part of the photo-isomerization follows a rotation-based mechanism\cite{Chang2004,Shao2008, Ootani2009}. 

In precending work we demonstrated that simple $\Delta$SCF-DFT calculations yield a qualitatively correct description of the involved excited-state PESs. They were shown to be in good agreement with higher level computations (CC2\cite{Christiansen1995}) and therefore provide a realistic representation of the mechanisms\cite{Maurer2011}. Figure \ref{refchange} reproduces these $\Delta$SCF curves (gray filled circles) following the ground-state optimized paths along the two main degrees of freedom, namely rotation around the central dihedral angle and inversion along one of the two central bending angles. To summarize the picture as arising from these simple $\Delta$SCF calculations, Z-Ab is predicted to be 0.58$\,$eV less stable than E-Ab, which compares nicely to the 0.6$\,$eV from experiment\cite{Schulze1977}. Both pathways show significant barriers in the ground state, whereas only in the rotational pathway a low lying state-crossing with the first excited state can be found at mid-rotation. The first 
excited state shows a minimum at mid-rotation and a very small barrier at mid-inversion. The second excited state, in fact corresponding to a number of $\pi\rightarrow\pi^*$ states, exibits two minima close to the positions of the ground state. These are on both pathways separated by sizable barriers. A state crossing close to the E-Ab equlibrium geometry with the S1 state can be found.

Figure \ref{refchange} also includes the data obtained when applying the le$\Delta$SCF approach as described in the previous section and using the ground-state orbitals of the isolated gas-phase molecule as reference orbitals at every position along the two pathways. Already from visual inspection it is possible to identify regions on both pathways where the difference to $\Delta$SCF is minimal and regions where the topology is not reproduced correctly. The assumption that the constraint states do not change significantly due to the excitation seems sufficiently justified very close to the equilibrium geometries, but fails at the transition-state geometries on the S2 state. In other words, in PES regions where ground-state orbitals are very good approximations to excited-state ones the difference is minimal. In contrast, in regions where due to excitation the orbital character and orbital ordering changes, effects can be quite large. In this respect, it is intriguing to note that the region of biggest error, 
namely the S2 state at mid-inversion, is also not correctly reproduced by lrTD-DFT when using ground-state PBE orbitals as a starting point\cite{Maurer2011}. Both effects have the same source, namely that GGA-DFT derived effective one-particle states are bad approximations to molecular resonances of the interacting many-particle system. This is especially true for virtual states\cite{Tozer1998a, DellaSala2002}.   

Also shown in Fig. \ref{refchange} are the curves calculated with le$\Delta$SCF when employing reference orbitals that were calculated with our le$\Delta$SCF$^{\text{EX}}$ approach. The corresponding results exactly reproduce the standard $\Delta$SCF curves, because they now include relaxation effects for all KS states. This nicely underscores the importance of including state relaxation in order to generate consistent-quality PESs. Having established the equivalence of the two methods for the gas-phase limit, we now proceed in the next section by analyzing the effects of a metal surface on the lowest-lying excited states of Azobenzene.

\subsection{Excited-state PESs of Azobenzene adsorbed on Ag(111) and Au(111)}
\label{PESchapter}

 \begin{figure*}
 \includegraphics[width=\linewidth]{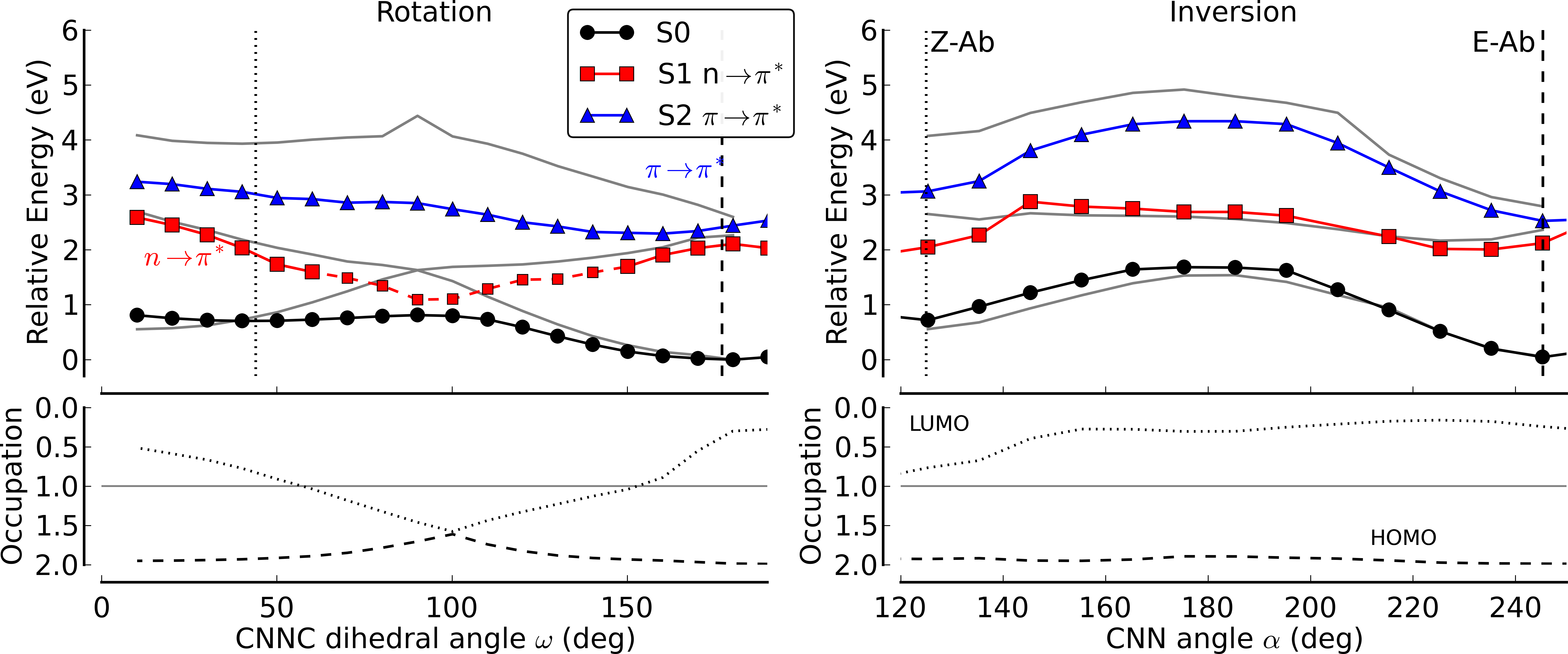}
 \caption{\label{PES-Ag} Upper panels: Minimum-energy paths of Azobenzene adsorbed on Ag(111) following rotation (left) or inversion (right). Shown are the ground-state energy (black), the first (S1, red) and second (S2, blue) excited states as well as the corresponding gas-phase potential energy curves (in gray). Regions marked with dashes are of increased inaccuracy due to methodological restrictions further outlined in the text. Vertical dashed and dotted lines on the sides depict the position of E-Ab and Z-Ab minima for the adsorbed molecule. Lower panels: For both degrees of freedom, rotation and inversion, the integrated occupation of the projected gas-phase HOMO (dashed line) and LUMO (dotted line) in the ground state are shown. The horizontal line marks half filling of an orbital.}
 \end{figure*}

 \begin{figure*}
 \includegraphics[width=\linewidth]{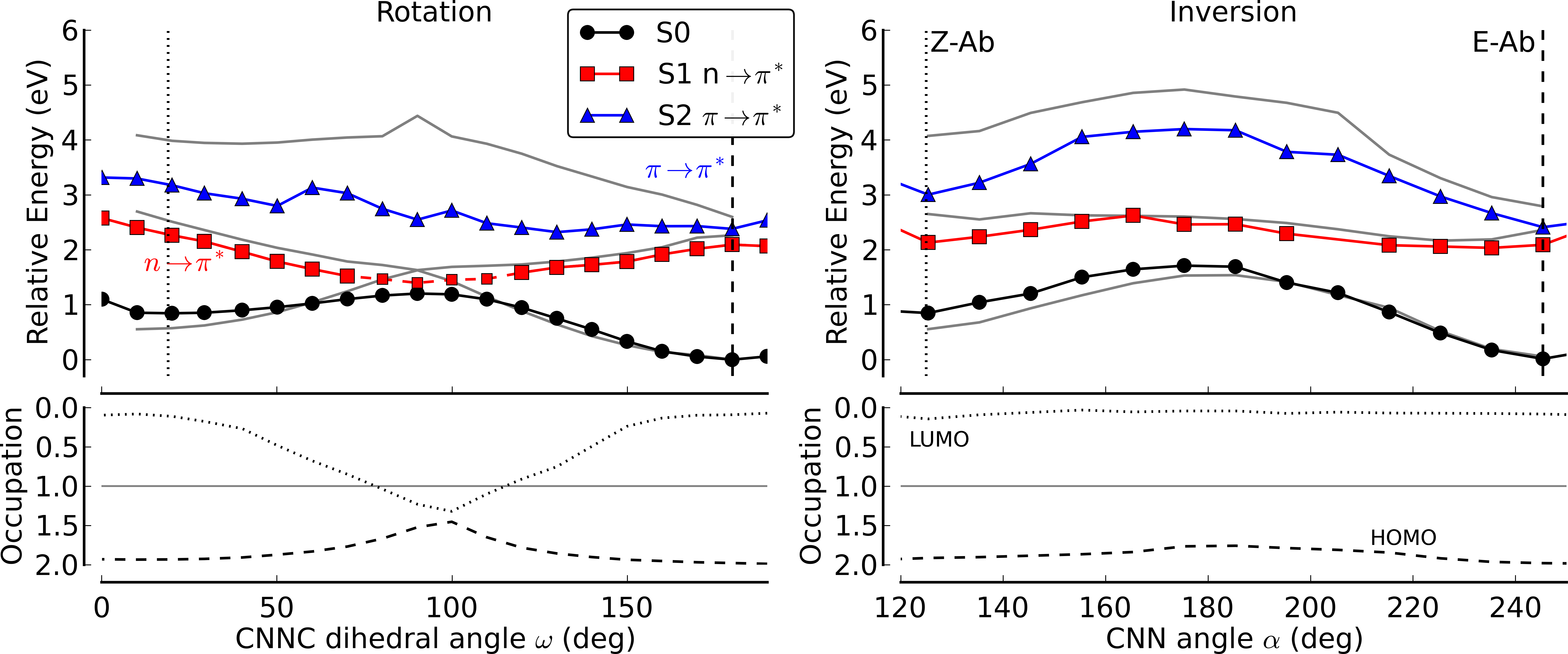}
 \caption{\label{PES-Au} Same as Fig. \ref{PES-Ag}, but for Azobenzene on Au(111).}
 \end{figure*}

Current knowledge of the mechanisms underlying photo-isomerization of Azobenzene and its derivatives on coinage metals is sparse. Photo-induced switching was hitherto only achieved for TBA on a Au(111) surface, with a significantly reduced cross section compared to the gas-phase or solvent case\cite{Hagen2007,Comstock2008}. The modified photo-absorbance of the adsorbed molecule led to the conclusion of a changed isomerization mechanism, where excitation happens indirectly via hole generation in the valence band and subsequent charge transfer from the molecule to the surface\cite{Wolf2009}. At Au(111) molecular switching was also achieved by resonant tunneling for Azobenzene \cite{Comstock2005,Alemani2008}, whereas to our knowledge no switching, neither light- nor current-induced, has been observed for Azobenzene or TBA on Ag(111). This lack of function on Ag(111) surfaces strongly supports recent DFT results pointing to an effective loss of bistability of both derivatives on Ag(111)\cite{Maurer2012}. 

Figures \ref{PES-Ag} and \ref{PES-Au} reproduce the corresponding ground-state paths of Azobenzene on Ag(111) and Au(111) following rotation and inversion. The barrier along inversion is almost unchanged, whereas the rotational barrier is strongly modified. Compared to the gas-phase case, the stability of the Z-Ab isomer is drastically reduced from a basin depth of 1$\,$eV to 0.05$\,$eV or 0.38$\,$eV at Ag(111) and Au(111), respectively (zero-point-energy corrected values from ref. \onlinecite{Maurer2012}). In the context of photo-induced E$\rightarrow$Z isomerization, this implies that vibrationally hot molecules on Ag(111) after deexcitation are liable to thermal re-isomerization to the E-Ab isomer. The bottom panels of Figs. \ref{PES-Ag} and \ref{PES-Au} show the integrated ground-state occupancies of the projected gas-phase reference orbitals corresponding to the HOMO and LUMO of Azobenzene. For E-Ab and Z-Ab as well as following geometries along the inversion degree of freedom no considerable charge is 
added to or withdrawn from these frontier orbitals on Au(111), cf. Fig. \ref{PES-Au} on the right. This indicates that the bonding in these molecular geometries is mainly physisorptive. Following the inversion isomerization of Azobenzene on Ag(111), cf. Fig. \ref{PES-Ag} on the right, we obtain a very similar picture, although the Z-Ab isomer already shows some charge transfer in the ground state. In contrast, following rotation we see that on both surfaces around mid-rotation the LUMO is more than half occupied and the HOMO loses considerable occupation. This is due to an orbital degeneracy of HOMO and LUMO at this point, which exists independent of metal surface adsorption. This leads to the formation of a strongly chemisorbed species at this point, further rationalising the ground state barrier reduction. The shift of the Z-Ab minimum towards higher $\omega$ angles (44$^\circ$) on Ag(111) together with the significant population of the LUMO creates a somewhat chemisorbed species in that case as well.

The le$\Delta$SCF$^{\text{EX}}$ method calculates the first and second excited states of these surface systems by adding an electron to the region of Hilbert space corresponding to the Ab gas-phase LUMO and removing an electron from HOMO or HOMO-1, respectively. The corresponding excited-state curves, cf. Figs. \ref{PES-Ag} and \ref{PES-Au}, for both degrees of freedom show very similar overall topologies compared to the respective gas-phase case. When following inversion on Ag(111) and Au(111), the S1 state is almost unchanged in comparison to gas-phase Azobenzene. A significant lowering of the excitation energy occurs only for geometries close to the Z-Ab minimum. For rotational isomerization, S1 state energies around mid-rotation are reduced simultaneously with the barrier reduction in the ground state, while excitations close to the equilibrium geometries are almost unchanged. The systematic downshift of the S2 state corresponds to a shift of about 1 eV on both coinage metal surfaces all along the 
pathway, except around the mainly physisorbed E-Ab geometry. Two very important features for the isomerization mechanism in gas phase are the state-crossings between S0 and S1 at mid-rotation and between S1 and S2 close to the E-Ab minimum. Both can, in principle, still be observed, suggesting that an intra-molecular isomerization mechanism analogous to the gas phase could also prevail at the surface. 

As most intriguing features of surface adsorption we thus see a stronger lowering of the S2 state compared to S1, and a stronger lowering of excitations for all geometries away from the E-Ab equilibrium structure. Both effects can be rationalized by the interaction of the molecular dipole with the image charge that is induced in the underlying metal substrate during adsorption. Azobenzene in the planar trans configuration shows no significant dipole orthogonal to the surface in the ground and both excited states. Yet, following the isomerization pathways towards the non-planar Z-Ab isomer, the z-component of the dipole in the ground state increases significantly to a gas-phase value of 3.0 Debye (D). The corresponding excited-state dipole moments for gas-phase Z-Ab are 2.3 D and 4.3 D for S1 and S2, respectively. The stronger polarisation of the S2 excited state thus leads to a stronger interaction of the molecular dipole with the image charge monopole and explains the particularly pronounced lowering of the 
S2 PES upon adsorption obtained in the le$\Delta$SCF$^{\text{EX}}$ calculations. An important point to mention here is that the variational treatment in le$\Delta$SCF and $\Delta$SCF approaches enables such an image charge build-up due to polarisation effects (opposed to lrTD-DFT treatments), although we emphasize that this is unlikely a quantitative account.

Another effect that modifies excited-state behaviour is the hybridization of molecular with surface states. A marker for the strength of hybridization is the change in ground-state occupation of the frontier orbitals, which we find much more pronounced for Ab adsorbed on Ag(111) than on Au(111). In regions where orbitals show occupancies very different from the gas phase, e.g. Z-Ab at Ag(111), we obtain PES changes that are more significant than for regions where occupancies do not change drastically. This effect is especially strong around mid-rotation, where the ground-state occupation of the LUMO already increases beyond one electron on both surfaces. This prohibits the full transfer of one further electron into the LUMO in the le$\Delta$SCF$^{\text{EX}}$ excited-state calculations and we instead only perform these calculations by enforcing a full two-electron occupation of the LUMO. In Figs. \ref{PES-Ag} and \ref{PES-Au} we mark these regions with dashed lines to emphasize the expected increased 
uncertainty due to the concomitant violation of the excitation constraint. We believe that these parts of the S1 curves can only serve as an upper estimate to the actual PES topology and attest that such situations of strong hybridization and charge transfer represent a clear limitation to the le$\Delta$SCF approach.

\begin{table}
\caption{\label{vertical} Vertical excitation energies (in eV) for the first Singlet excited state of Z-Ab (S1) and the second Singlet excited state of E-Ab (S2), as well as the difference between them. Reference values shown are taken from experiments in solvent \cite{Rau2003, Andersson1982} and from high-level quantum chemical (RI-CC2) calculations for the isolated molecule \cite{Maurer2011}.}
\begin{center}
\begin{tabular}{cccc}
\hline\hline  & Z-Ab S1 & E-Ab S2 & Energy Difference \\ \hline
Exp.$^a$    &   2.87/2.92   &  3.89/4.12 & 1.02/1.2  \\ 
CC2   @gas$^b$              & 3.00 &    4.07    & 1.07 \\
$\Delta$SCF-B3LYP @gas$^b$   & 2.30 & 3.33 & 1.03  \\
$\Delta$SCF-PBE   @gas$^b$   & 2.10 & 2.98 & 0.88  \\
le$\Delta$SCF-PBE @Ag(111)   & 2.03 & 2.44 & 0.41  \\
le$\Delta$SCF-PBE @Au(111)   & 2.27 & 2.38 & 0.11  \\ \hline\hline
$^a$: refs. \onlinecite{Rau2003, Andersson1982}, $^b$: ref. \onlinecite{Maurer2011}
\end{tabular}\end{center}
\end{table}

Notwithstanding, even when only taking them qualitatively, the obtained results clearly show that an explicit treatment of hybridization, charge transfer and image charge effects is necessary to appropriately describe ground- and excited-state PESs of a functional molecule like Azobenzene when adsorbed at metal surfaces. Investigating Ab in a van-der-Waals potential to merely mimick the effect of surface-modified molecular geometries on the switching function, Flo{\ss} \textit{et al.} \cite{Floss2012} recently reported only a small increase in conversion times and decrease of photo-yield compared to the gas phase, while otherwise the photo-isomerization was unaffected. Without yet embarking on actual dynamical simulations, the le$\Delta$SCF-obtained PESs topologies already indicate that much larger effects are induced by the metal electronic structure. Notably this is the image charge induced lowering of excited-state PESs, which due to the varying degree of state polarization and molecular dipole moment 
does not occur globally, but differentially "skews" individual state topologies and vertical excitation energies. For the present system this leads to a strong lowering of the S2 state particularly around the Z-Ab geometry. 

By itself this image-charge lowering might already rationalize a significantly reduced switching efficiency of the traditional intra-molecular gas-phase isomerization mechanism at the surface: This mechanism proceeds via initial excitation to S2 and fast population transfer to S1 for the E$\rightarrow$Z isomerization, as transition to S1 is symmetry forbidden in the E-Ab geometry. The back-reaction Z$\rightarrow$E instead involves direct excitation to S1. In the gas phase the corresponding vertical excitation energies for the two reactions differ substantially (cf. Table \ref{vertical}) and allow the two isomerizations to be selectively induced by light with two largely differing wave lengths. In contrast, at the surface our le$\Delta$SCF$^{\text{EX}}$ results suggest that the selective image-charge induced S2 lowering reduces this difference for the two transitions substantially. While in the gas phase it amounts to more than 1\,eV, particularly at Au(111) the difference between E-Ab S2 and Z-Ab S1 reduces 
to 0.11 eV. Considering an additional state broadening at the surface, this proximity of the two different excitations alone might then already cause a significant loss of switching efficiency via this intra-molecular mechanism as the forward and backward isomerization can simply no longer be selectively triggered.

At least qualitatively, these findings should also be robust against the other clear limitation of le$\Delta$SCF, namely the one imposed by the employed approximate DFT functional. Already in the gas phase our preceding work demonstrated that GGA-PBE based $\Delta$SCF (but also GGA-PBE based lrTD-DFT) severely underestimated absolute vertical excitation energies for Azobenzene compared to accurate quantum-chemical (RI-CC2) calculations \cite{Maurer2011}. These were primarily global shifts of entire respective excited-state PESs though and largely left topological features like barriers unchanged. Addition of exact exchange can remedy these self-interaction induced shortcomings of the semi-local functional for gas-phase Azobenzene \cite{Maurer2011}. However, simultaneously it would remove much of the balanced error cancellation in the description of the metal substrate\cite{Hu2007,Stroppa2008}. For metal-surface adsorption there is at present no feasible and equally efficient alternative to semi-local DFT. 
GGA-PBE based le$\Delta$SCF excited-state PESs thus have to be seen in light of the self-interaction induced overpolarizability and wrong relative positions of molecular and substrate states, which will affect the observed image charge and hybridization effects upon adsorption. While thus certainly not quantitative, the approach still enables in our view an effective first account of the electronic structure and charge distributions at the metal surface and is thus a viable means to generate further insight into the intricacies of surface functionality of large organic molecules like Azobenzene.

\section{Conclusions}
\label{conclusion}

We presented an alternative implementation of the le$\Delta$SCF method of Gavnholt {\em et al.} \cite{Gavnholt2008} and necessary modifications to allow its application to complex metal-surface adsorbed chemical reactions. The current method provides a computationally efficient way to describe low-lying localized excited states in large periodic systems. The correct calculation of reference states that are used to generate the resonances assures consistent quality of excited-state potential energy surfaces and also sets the connection to simple $\Delta$SCF in the limit of vanishing hybridization, in the surface context between adsorbate and metal substrate. For the example of photo-induced isomerization of Azobenzene at (111) coinage metal surfaces we illustrated that the approach yields an account of the additional stabilization due the interaction of large excited-state dipoles with the substrate image charge and hybridization-induced state renormalization. As such the method at least qualitatively 
describes most important physical effects that arise from the interaction with the electronic structure and charge distributions at the metal surface, and thus allows to discuss surface effects on the molecular functionality beyond the level of surface-modified adsorbate geometries. For the Azobenzene showcase system this is already highlighted by the observed unbalanced shifts of the intra-molecular excited-state energies of E-Ab and Z-Ab, which leads to an alignment of isomerization-inducing resonances. Reducing the ability to selectively trigger back and forth isomerization, this could be a first important piece in the puzzle to understand the strongly reduced isomerization efficiency at the surface - at least of the traditional gas-phase mechanism. 

We believe that the approach presented in this work, although approximate in nature, enables a semi-quantitative account of excited-state properties for large-scale systems and might prove to be very useful specifically for large hybrid organic/metallic interfaces. While it may never replace theoretically rigorous methods, such as TD-DFT or many-body perturbation theory, it fills a gap in the current methodological spectrum, where these more accurate methods are not yet applicable due to their computational expense or where currently used approximations in the xc-kernel or self-energy description in these methods cause a lack of consistent accuracy. Independent from the development of these schemes there will always be the need for very efficient treatments that allow fast screening on a qualitative or, when solid benchmarking is done, possibly semi-quantitative level. 

\begin{acknowledgments}
Funding through the German Research Council is gratefully acknowledged. RJM would like to thank Thomas Olsen for helpful comments. Computer resources for this project have been provided by the Gauss Centre for Supercomputing/Leibniz Supercomputing Centre under grant id: pr63ya.
\end{acknowledgments}


%

\end{document}